# Wafer-Scale Assembly of Semiconductor Nanowire Arrays by Contact Printing

Zhiyong Fan, Johnny C. Ho, Zachery A. Jacobson, Roie Yerushalmi, Robert L. Alley, Haleh Razavi, Ali Javey*

Department of Electrical Engineering and Computer Sciences, University of California at Berkeley, Berkeley, CA 94720.

Materials Sciences Division, Lawrence Berkeley National Laboratory, Berkeley, CA 94720

* **Correspondence to**: ALI JAVEY ajavey@eecs.berkeley.edu

## Abstract

Controlled and uniform assembly of "bottom-up" nanowire (NW) materials with high scalability has been one of the significant bottleneck challenges facing the potential integration of nanowires for both nano and macro electronic circuit applications. Many efforts have focused on tackling this challenge, and while significant progress has been made, still most presented approaches lack either the desired controllability in the positioning of nanowires or the needed uniformity over large scales. Here, we demonstrate wafer-scale assembly of highly ordered, dense, and regular arrays of NWs with high uniformity and reproducibility through a simple contact printing process. We demonstrate contact printing as a versatile strategy for direct transfer and controlled positioning of various NW materials into complex structural configurations on substrates. The assembled NW pitch is shown to be readily modulated through the surface chemical treatment of the receiver substrate, with the highest density approaching ~8 NW/μm, ~95% directional alignment and wafer-scale uniformity. Furthermore, we demonstrate that our printing approach enables large-scale integration of NW arrays for various device structures on both Si and plastic substrates, with a controlled semiconductor channel width, and therefore ON current, ranging from a single NW (~10 nm) and up to ~250 μm, consisting of a parallel array of over 1,250 NWs.

In recent years, synthetic nanomaterials, such as carbon nanotubes and semiconductor nanowires, have been actively explored as the potential building blocks for various electronic applications as they offer advanced properties arising from their miniaturized dimensions[1-11]. For instance, they may enable further scaling of the devices down to the molecular regime with enhanced performances [2-7, 10, 11] while enabling new functionalities and therefore leading the way to novel technological applications [1, 8, 9, 12-15] However, inarguably, one of the most significant challenges still facing the bottom-up approach is the large scale assembly of the nanomaterials at well defined locations on substrates with high reproducibility and uniformity. Many efforts have focused on tackling the controlled assembly [12, 13, 16-26], and while significant progress has been made, it still remains an unresolved obstacle. Here, we present a novel strategy for wafer scale assembly of highly ordered and aligned NW arrays on substrates with controlled pitch and density for electronic applications. The method is based on a simple contact printing process that enables the direct transfer and positioning of NWs from a donor substrate to a receiver chip. The process is compatible with a wide range of receiver substrates, including Si and flexible plastics. The dynamics and the mechanism of this transfer process are also explored through a series of systematic studies that has enabled us to gain further control over the NW assembly.

Our contact printing method involves directional sliding of a growth substrate, consisting of a dense "lawn" of NWs, on top of a receiver substrate coated with lithographically patterned resist (Figure 1a). During the process, NWs are in effect combed by the sliding step, and are eventually detached from the donor substrate as they are anchored by the van der Waals interactions with the surface of the receiver substrate; therefore, resulting in the direct transfer of aligned wires to the receiver chip with un-transferred NWs remaining unbroken on the donor substrate according to scanning electron microscopy (SEM) observation. An important aspect of our strategy, as compared to our previously reported dry transfer method [13], is the use of octane/mineral oil (2:1 v:v) mixture as a lubricant during the transfer process in order to significantly reduce the NW-NW friction while sliding the two substrates. Such friction enhances breaking and grinding of NWs hampering the alignment. On the other hand, the lubricant does not affect significantly the interactions of the chemically modified polar surface with the sliding NWs due to the low dielectric constant ($\varepsilon \sim 2$). This is further supported by our observations where essentially no NWs were transferred to the receiver substrate for similar printing conditions using aqueous detergent solution ($\varepsilon \sim 80$) for lubrication rather than oil (not shown). Maintaining NW-substrate interactions is essential for the assembly process where dynamic friction is a key for alignment of NWs and the detachment of NWs from the donor substrate. After NW printing, the spacer resist is removed by a standard lift-off process using acetone, leaving behind the nanowires that were directly transferred to the patterned regions of the substrate. The process is highly generic for a wide range of NW materials, including Si, Ge, and compound semiconductors, and for the entire NW diameter range, d=10-100 nm, that was explored in this study.

The optical and SEM images of the assembled GeNWs on a Si/SiO$_2$ (50 nm, thermally grown) receiver substrate are shown in Figs 1b-c, clearly demonstrating the uniformity of the well-aligned NW films. Notably, the transferred NWs form a single layer without any significant multilayer stacking (Fig. 1c) which is attributed to the weak interactions between the unmodified NWs. This rationale is further supported by the SEM observation showing no clustering of NWs on donor or receiver substrates. Therefore, the printing process is self-limiting with the transfer of NWs restricted to a monolayer. A self-limiting directional assembly process is highly desirable for various electronic applications where uncontrolled vertical stacking of the wires can result in reduced gate coupling, and therefore, switching performances. In order to achieve controlled stacking of the wires, a multistep printing methodology was developed as demonstrated in Figures 1d&e where Si nanowire crosses are enabled through a two step process. First, a layer of SiNW parallel arrays was printed on a receiver substrate. The sample was then spin-coated with a thin film of polymethylmethacrylate (PMMA, ~40 nm thick) or a photoresist (Shipley, ~40 nm) to serve as a buffer layer, followed by a second printing step normal to the direction of the first layer. Finally, the polymer buffer layer was etched away by a mild $O_2$ plasma process (60 W, 2 min), therefore, resulting in the assembly of large arrays of NW crosses. The printed NW crosses further demonstrate the versatility of the contact printing methodology presented here for hierarchical assembly of the bottom-up nanowires into a variety of geometric structures with high complexity and for a wide range of potential applications. For instance, the controlled assembly of the NW crosses may enable novel optoelectronic applications via heterogeneous integration of *p*- and *n*-type NW materials [27, 28].

To quantitatively characterize the printed NW films, statistical analyses were conducted for the length and degree of the orientation alignment of the printed NWs. We observed that the average length of the printed NWs linearly increases with the length of the donor substrate nanowires (Fig. 1f), approaching ~40 μm long for donor NW length of ~80 μm which highly depends on resist thickness and pressure [29]. Notably, we achieve high degree of alignment in the printed NW films. As shown in Figure 1g, >90% of NWs are highly aligned in the direction of sliding without any significant sensitivity to the NW length. A NW was considered to be misaligned if its axis formed an angle >5 º with respect to the sliding direction.

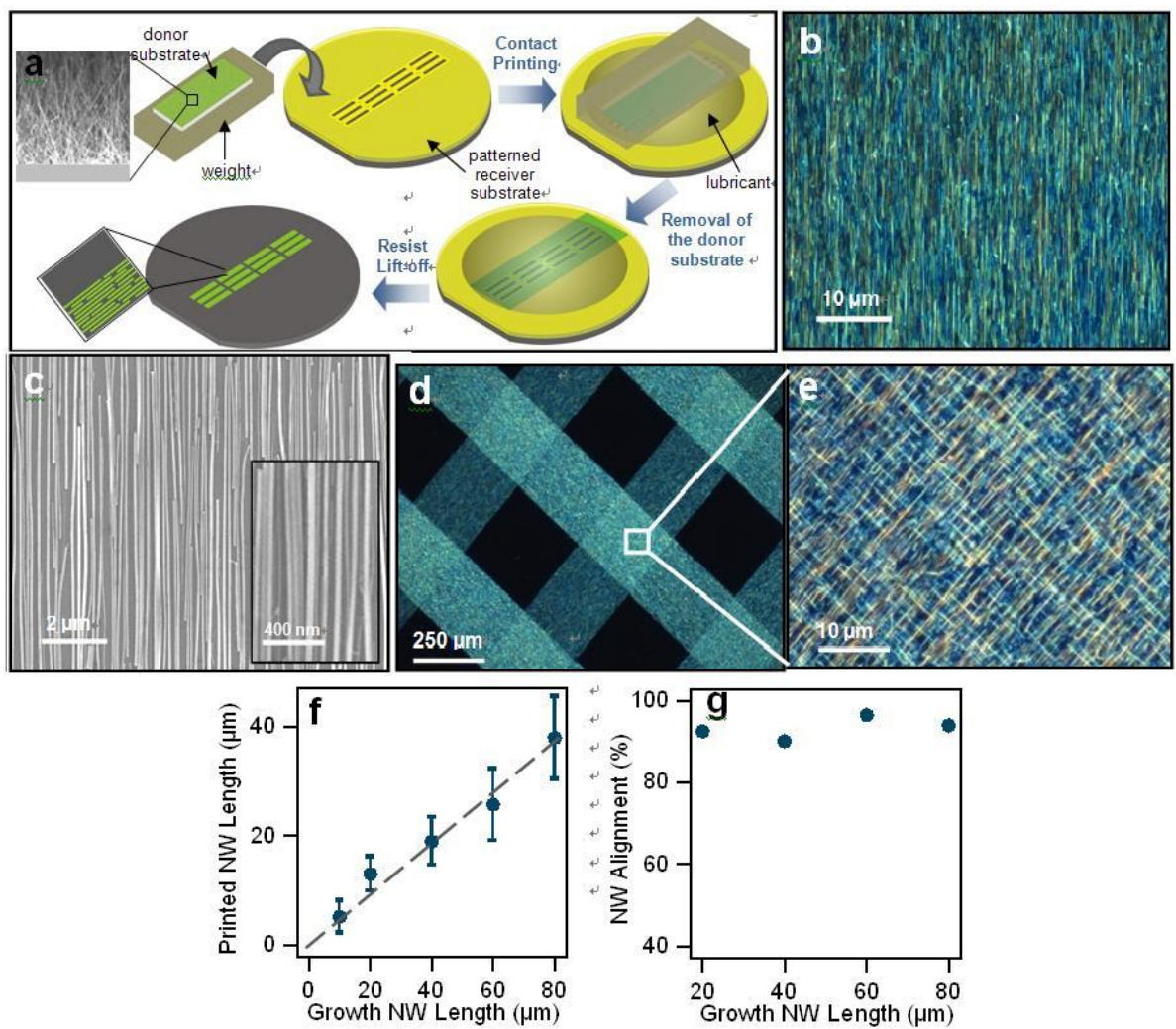

**Figure 1. a**, Schematic of the process flow for contact printing of nanowire arrays. **b**, Dark-field optical and **c**, SEM images of GeNWs printed on a Si/SiO$_2$ substrate showing highly dense and aligned monolayer of nanowires. The self-limiting process limits the transfer of NWs to a monolayer. **d**, and **e**, optical images of double layer printing for SiNW cross assembly. **f**, Average length of the printed nanowires as function of their original length on donor substrate, showing a linear trend with a slope of ~0.5. **g**, The percentage of the printed nanowires aligned on the receiver substrate for various NW length scales. A nanowire is considered misaligned if its axis forms an angle > 5 º with respect to the printing direction.

To shed light on the transfer mechanism and the process dynamics, and to gain further control over the printing process, we have explored the role of surface chemical modification of the receiver substrate on the density of the printed NWs. Surface modification of $Si/SiO_2$ (50 nm) substrates was achieved by well-established siloxane-based condensation chemistry of various compounds including (heptadecafluoro-1,1,2,2-tetrahydrodecyl) dimethylchlorosilane (Gelest, Inc.), N-trimethoxysilylpropyl-N,N,N-trimethylammonium chloride (Gelest, Inc.), and 3-Triethoxysilylpropylamine (Sigma-Aldrich, Inc.) to define $-CF_3$, $-N(Me)_3^+$, or $-NH_2$ terminated surfaces, [9, 30] respectively. To carry out the reactions, solutions of ~0.5 vol% of the above compounds were prepared with either hexane or ethanol as the solvent. $Si/SiO_2$ substrates were then reacted with the respective solutions for 45 min, followed by a through wash with the solvent and baking at 120°C for 20 min. For the poly-L-lysine functionalization, the substrate was coated with ~0.1% w/v of poly-L-lysine in water (Ted Pella, Inc.) for 5 min, and was then thoroughly rinsed with DI water. Notably, the poly-L-lysine functionalization can be carried out directly on top of the photolithographically patterned substrate since it does not react with the Shipley 1805 photoresist.

Figure 2 shows the printed density of as-grown GeNWs (chemically unmodified, d=15 nm) on chemically modified $SiO_2$ receiver substrates. For the $-CF_3$ terminated monolayers, we observed almost no transfer of nanowires ($<10^{-3}$ NW/μm) from the donor to the receiver chip while an identical printing process on $-NH_2$ terminated monolayers resulted in high density of transferred nanowires, approaching ~ 8 NW/μm. Such a major density modulation of ~4 orders of magnitude demonstrates the key role of the surface interactions on the printing outcome. $CF_3$ functionalized surfaces are well known to be highly hydrophobic and "non-sticky", therefore, minimizing the adhesion of NWs to the receiver substrate during the sliding process. As a result, the wires remain rather unbroken on the donor substrate without being transferred to the receiver chip. On the other hand, -NH$_2$ terminated surfaces interact effectively with the NW surface [20] to yield a high density transfer through strong bonding interactions. This demonstrates that our transfer process requires strong NW to receiver substrate interactions, which eventually results in the breakage of the wires and direct transfer to the receiver substrate during the sliding step. Notably, when the contact printing is carried out without the use of lubricants, dramatically lower density contrast is observed for the various surface treatments, which hints that the use of lubrication is essential for gentle and controlled transfer of NWs by lowering non-selective mechanical friction.

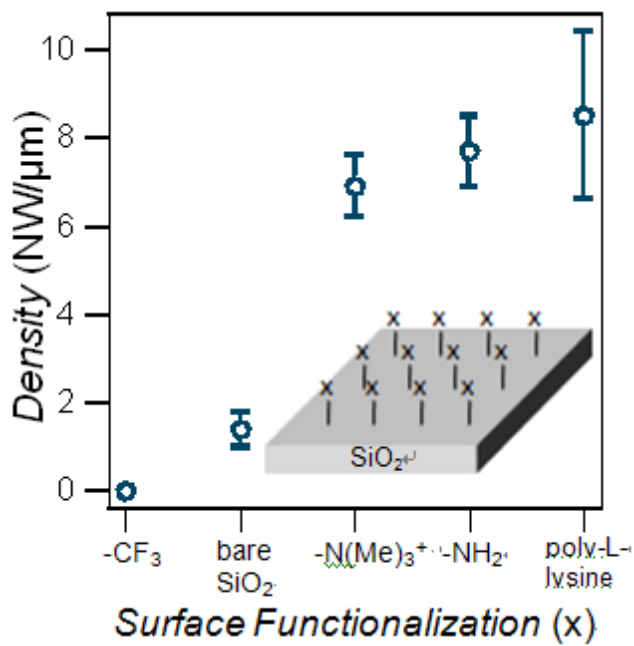

**Figure 2.** Nanowire printing density modulation by receiver substrate surface functionalization. GeNWs (d~15 nm) were used for this study with a Si/SiO$_2$ receiver substrate. Bare SiO$_2$ corresponds to using the untreated substrate while –CF$_3$, and –N(Me)$_3^+$, and -NH$_2$ correspond to the surface modified functional groups. Note that, as expected, poly-L-lysine functionalization results in a larger standard deviation due to the less uniform coverage of the surface by polymer thin film deposition rather than monolayer formation.

The potency of our contact printing process is further demonstrated through successful assembly of regular arrays of single NWs at predefined locations on the substrate. To precisely control the location of single nanowires during the printing, arrays of fine patterns with ~200 nm width and 2 μm pitch (Fig. 3a inset) were defined on a PMMA resist with electron beam lithography (JEOL 6400, NPGS). The patterned substrate was then functionalized with amine-terminated monolayer to increase its adhesion to NWs. The followed GeNW printing and PMMA lift-off in acetone yielded a regular array of highly ordered single NWs with a 2 μm period as depicted in Figures 3a&b. Interestingly, the patterned width of ~200 nm that is required to achieve single NW trapping corresponds well to the average NW pitch of ~130 nm for the same NW material and surface treatment (Fig. 2). Importantly, our contact printing process is highly scalable. We have been able to successfully assemble highly aligned and dense (~7 NW/μm) nanowire parallel arrays on a 4" Si wafer (with 50 nm $SiO_2$ functionalized with amine-terminated monolayer) as shown in Figure 3c. Interestingly, the parallel array NW films, appearing as gray strips in the optical photograph, show a very uniform color contrast across the entire wafer which is a strong indicator of the uniform density and alignment of the assembled NWs. Notably, the NW film density that is attained from our contact printing method is higher by ~ 20x over that of the other scalable assembly methodologies, such as the bubble blown technique [17]. A high NW density is desirable for achieving dense arrays of single NW devices, or high ON currents for the case of thin film transistors with parallel array NW channels.

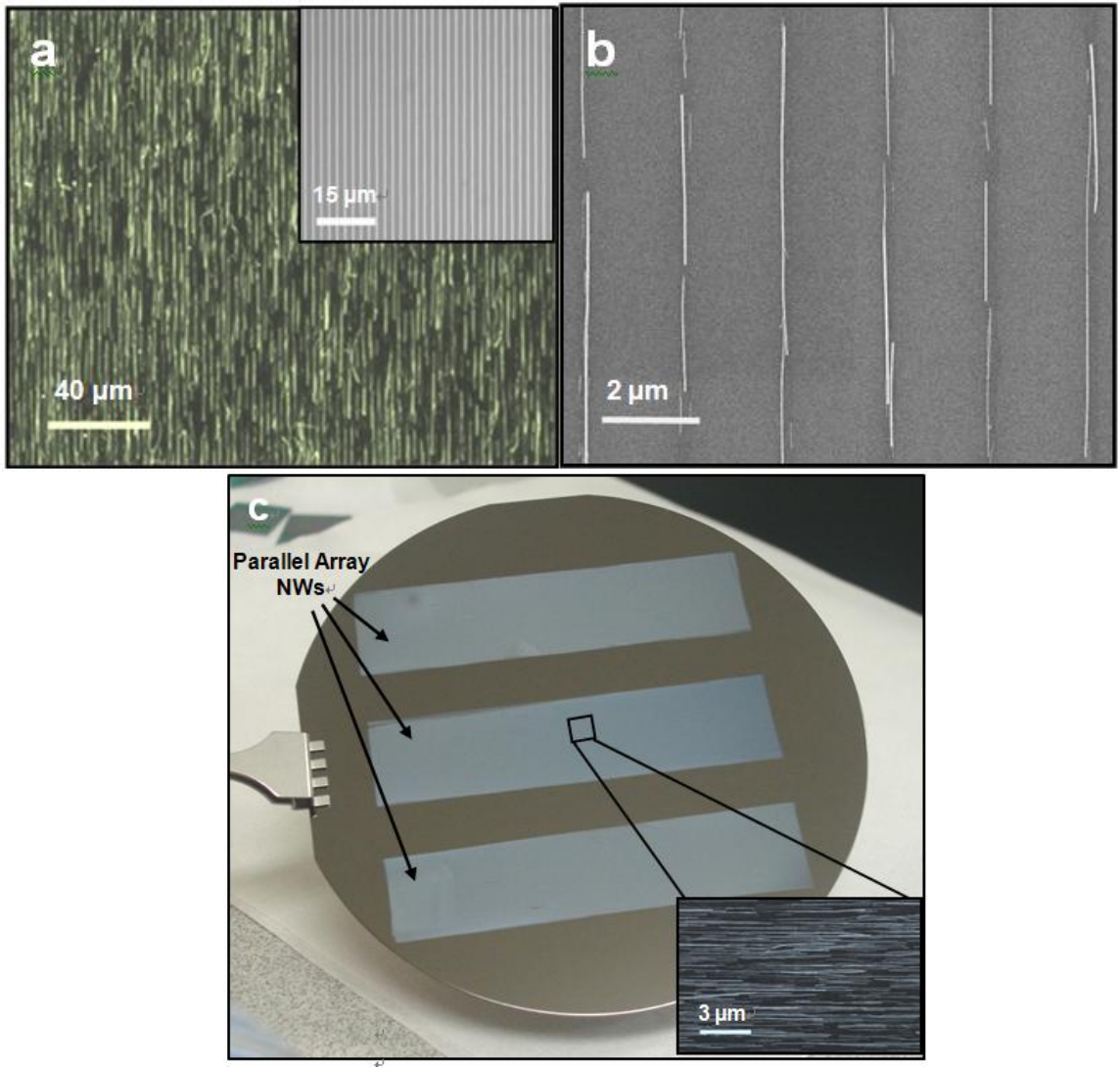

**Figure 3.** Single nanowire printing. **a**, Optical and **b**, SEM images showing regular array assembly of single Ge nanowires at predefined locations on a Si/SiO2 substrate. The inset in **a** shows an optical image of the e-beam patterned receiver substrate prior to the contact printing, consisting of an array of 200 nm wide patterns on a PMMA resist. **c**, Large area and highly uniform parallel arrays of aligned GeNWs were assembled on a 4" Si/SiO$_2$ wafer by contact printing. The inset is an SEM image of the printed NW film, showing a density ~ 7 NW/μm.

The ability to uniformly assemble both single and highly dense parallel arrays of NWs on substrates and in large-scale leads the way for exploring a wide range of electronic device structures based on NWs. This is of paramount importance since various applications have different criteria for the desired device characteristics, for example some demanding high density while others requiring high current drives. In order to explore the feasibility of our contact printing strategy for various electronic structures, we have studied the effect of channel width scaling on the device performance and characteristics. High mobility core/shell Ge/Si (~15/5 nm) NWs [13, 31, 32] were printed into regular arrays and configured as back-gated FETs by defining Ni/Pd (5/45 nm) source (S) and drain (D) contacts with channel widths ranging from a single NW (~20 nm) up to ~250 µm (Figure 4a) and channel length of ~ 2 µm. The average ON current of NW FETs as a function of the channel width, and therefore the number of NWs per channel, is shown in Figure 4 b. It can be clearly seen that the ON current linearly scales with the channel width with a slope of ~ 5 µA/µm, corresponding to ~ 5 NW/µm since on average a single Ge/Si NW in an un-optimized back-gated geometry delivers ~ 1 µA. This NW density is consistent with the expected NW pitch of similar diameter and receiver substrate surface functionalization as demonstrated in Figure 2. It is worthy to note that in the future, the device performance can be drastically improved by ~1-2 orders of magnitude through channel length scaling, and high-κ dielectric and metal top gate integration to enable better electrostatic modulation of the NWs as previously demonstrated for the case of single Ge/Si NW FETs [32]. The highly linear scaling of the ON current with the channel width demonstrates the uniformity and reproducibility of the well aligned NW arrays that are enabled through our contacting printing process.

A unique aspect of our printing process is its compatibility with various substrate materials, including plastics. In fact, we have successfully printed *p*-SiNW arrays on flexible plastic substrates (Kapton, DuPont) and configured them into novel diode structures by using asymmetric Pd-Al S-D contacts. High work function Pd metal forms a near ohmic contact to the valence band of the *p*-SiNWs whereas low work function Al results in a Schottky contact. Figure 4c shows an optical photograph and schematic of the parallel array NW diodes that were fabricated on Kapton. Notably, no doping profiling (for example, *p-n* junctions) of NWs is used for obtaining the diodes. This is critical since such a doping process would require high temperature diffusion and/or activation of the dopants which is not compatible with the plastic substrates. The *I-V* characteristic of a representative NW Schottky diode is shown in Figure 4d, demonstrating clear rectifying effect with four orders of magnitude difference in the current for the reverse versus forward bias condition. The ability to readily obtain diode arrays on plastics is of significant interest for a wide range of potential optoelectronic applications, including solar cells, light emitting diodes and light sensors.

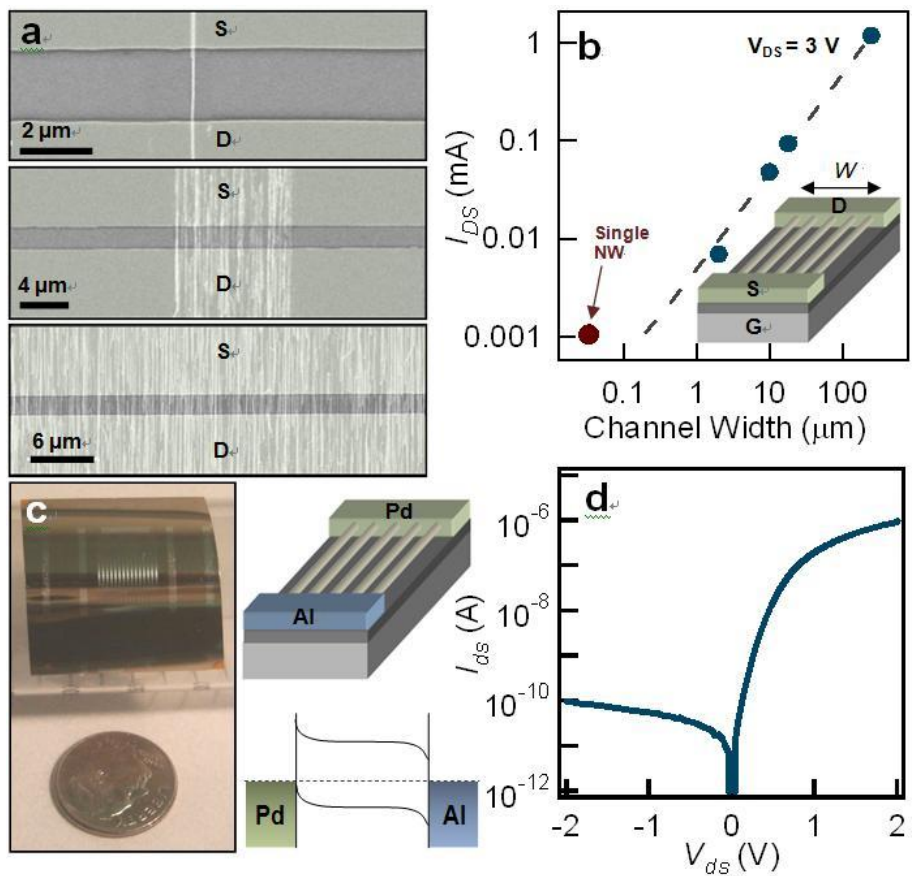

**Figure 4.** Devices based on printed nanowire arrays. **a**, From top to bottom SEM images of back-gated single NW FET, 10 μm and 250 μm wide parallel arrayed NW FETs. High mobility Ge/Si (15/5 nm) NWs were used as the channel material. **b**, ON current as a function of channel width scaling, showing a highly linear trend. **c**, Optical photograph and schematic of a novel diode structure fabricated on parallel arrays of *p*-SiNWs on a flexible plastic substrate. Asymmetric Pd-Al contacts are used to obtain Schottky diodes with the Pd forming near ohmic contact to the valence band and Al resulting in a Schottky interface. **d**, *I-V* characteristic of a representative SiNW Schottky diode with a channel width ~ 200 μm and length ~ 2 μm.

In summary, a generic approach was developed to directly transfer regular arrays of semiconductor NWs from the growth to patterned receiver substrates. The transferred NWs are highly ordered and aligned, enabling the assembly of both single NWs and densely-packed parallel arrays of NWs with high uniformity across an entire wafer. In addition, various device structures, such as field-effect transistors on single and parallel arrays of NWs and Schottky diodes with asymmetric contacts were fabricated on rigid and flexible substrates, which demonstrates the potency and versatility of our method for integrating NW materials for electronic and optoelectronic applications.

**Acknowledgements:**

This work was financially supported by MARCO/MSD Focus Center Research Program, Lawrence Berkeley National Laboratory, and a Human Frontiers Science Program fellowship (R.Y.).